\documentclass[letterpaper,10pt,conference]{ieeeconf}
\IEEEoverridecommandlockouts
\usepackage{graphicx}
\usepackage{xcolor}
\usepackage{booktabs}
\usepackage{amsmath,amssymb}
\usepackage{multirow}
\usepackage{cite}
\usepackage[hidelinks]{hyperref}
\usepackage[nameinlink,noabbrev]{cleveref}

\title{Mask2Restore: Self-Supervised Ultrasound Despeckling via Inpainting}
\author{Xuesong Li$^{a,b}$, Yingtai Xu$^a$, Zhongliang Jiang$^c$, Nassir Navab$^{a,b}$, and Yuan Bi$^{a,b}$\thanks{$^a$ Computer Aided Medical Procedures (CAMP), Technical University of Munich, Munich, Germany; $^b$ Munich Center for Machine Learning (MCML), Munich, Germany; $^c$ Department of Mechanical Engineering, The University of Hong Kong, Hong Kong SAR, China.}}

\begin{document}
\maketitle
\begin{abstract}
Medical ultrasound (US) is inherently degraded by speckle, a granular interference pattern that is often treated as a complex form of noise in image restoration. However, unlike random noise, US speckle originates from coherent scattering within tissue and is therefore highly spatially dependent and deterministic under fixed acquisition conditions, making US speckle suppression fundamentally different from natural image denoising. Because speckle-free US targets are unavailable in practice, self-supervised denoising is necessary. Blind-spot networks (BSN) are the dominant self-supervised paradigm for natural images, but their pixel-wise masking strategy assumes spatially independent noise, an assumption poorly matched to US speckle, which is spatially correlated over multiple pixels rather than pixel-wise independent. To address this mismatch, we propose Mask2Restore, a self-supervised US despeckling framework that reformulates despeckling as contextual inpainting with block-wise masking on single noisy images. Unlike pixel-wise BSN masking, block-wise masking addresses this multi-pixel speckle correlation by removing locally correlated speckle neighborhoods and shifting the reconstruction cues used by the network from adjacent speckle correlations to broader anatomical context. We further introduce cross-resolution context regularization (CRCR), which suppresses residual speckle bias by enforcing consistency across multi-resolution predictions. Experiments on simulated and in vivo carotid US, unseen fine-structure cases, and downstream cardiac segmentation demonstrate improved speckle-detail trade-offs, better preservation of fine anatomical structures, and practical value for subsequent image analysis.

\end{abstract}

\vspace{0.3em}

\section{Introduction}
\label{sec:intro}

\begin{figure*}[t]
\centering
\includegraphics[width=0.8\textwidth]{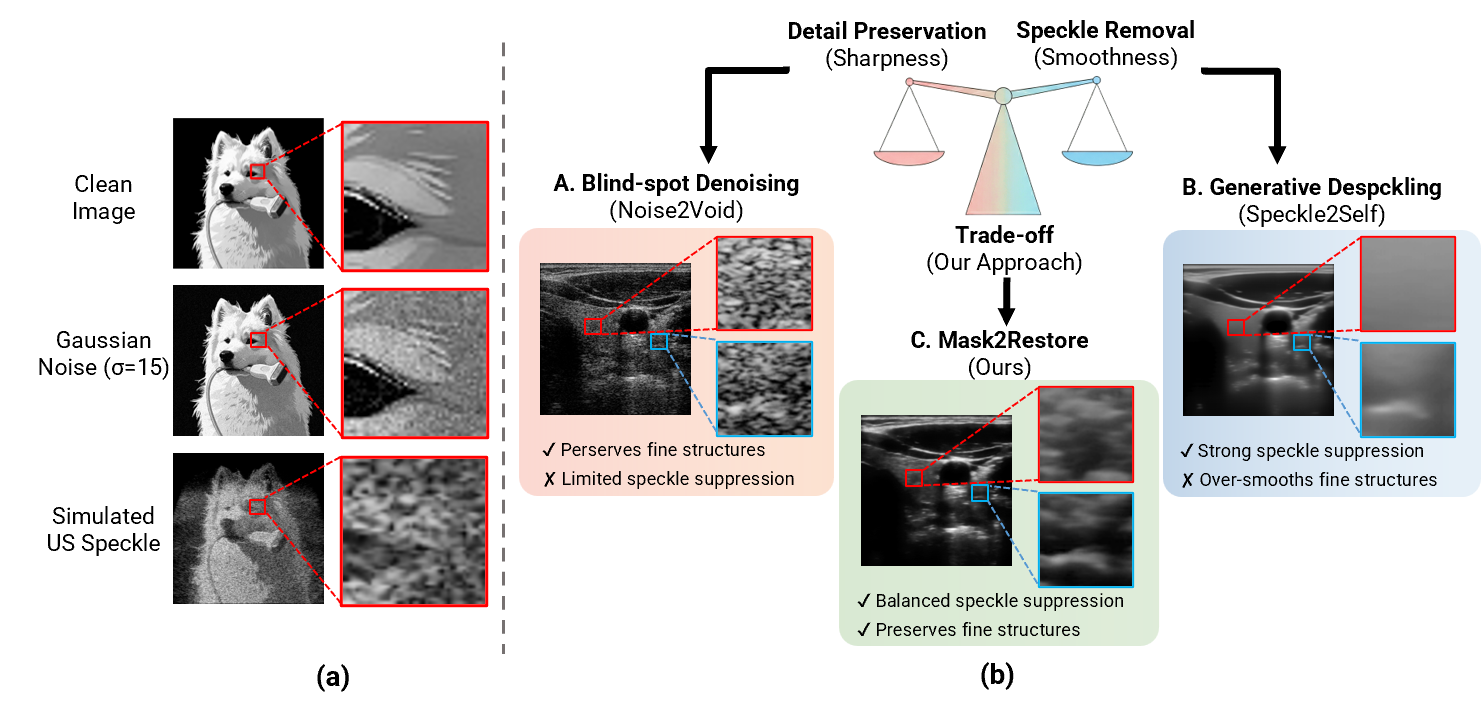}
\caption{\textbf{(a)} Comparison between pixel-wise Gaussian noise ($\sigma=15$) and simulator-generated US speckle. Unlike random Gaussian noise, US speckle forms spatially correlated granular patterns that can obscure fine image structures. \textbf{(b)} Paradigm-level illustration of the speckle-detail trade-off in self-supervised US despeckling, where A denotes blind-spot denoising, B denotes generative despeckling, and C denotes Mask2Restore. Methods that prioritize structural fidelity may leave residual speckle, whereas stronger despeckling priors can over-smooth fine anatomy. Mask2Restore aims to improve this trade-off by suppressing speckle while preserving fine anatomical structures.}
\label{fig:denoising_typesPLUSnoisetypes}
\end{figure*}

Medical ultrasound (US) is widely used in clinical practice due to its low cost, real-time capability, and safety~\cite{jiang2023robotic,li2025speckle2self,li2025semantic}. However, its image quality is inherently limited by US speckle, a granular interference pattern arising from echoes scattered by tissue microstructures~\cite{szabo2013diagnostic}. US speckle substantially degrades image clarity and contrast, thereby complicating both clinical interpretation and downstream computer-aided analysis~\cite{szabo2013diagnostic,jiang2023robotic}.
Unlike random noise such as Gaussian or Poisson noise, US speckle exhibits strong spatial dependency and a clustered multi-pixel appearance. As illustrated in \cref{fig:denoising_typesPLUSnoisetypes}(a), simulator-generated US speckle produces spatially correlated granular distortions that can obscure fine structures, so suppressing speckle while preserving details remains a key challenge.

Classical despeckling methods (e.g., filtering, anisotropic diffusion, and non-local means variants) can be effective but often require careful parameter tuning and may introduce additional artifacts or oversmooth fine structures~\cite{zhang2015wavelet,krissian2007oriented,coupe2009nonlocal,zhu2017non,zhou2019iterative,jiang2023skeleton,jiang2024class}.
Recent advances in fully supervised deep learning for natural images, particularly convolutional neural networks (CNNs), have enabled innovative data-driven approaches that learn direct mappings from corrupted inputs to clean targets without explicit noise modeling~\cite{zhang2017beyond}. However, as in many medical imaging modalities, clean targets for US speckle suppression are generally unavailable~\cite{huang2021neighbor2neighbor,pang2021recorrupted,li2025speckle2self}.
In modalities such as computed tomography (CT), pseudo-clean targets can be approximated by increasing the radiation dose under controlled conditions~\cite{wang2023ctformer}.
By contrast, in US imaging, speckle is inherent to the imaging process, making truly speckle-free acquisitions infeasible in practice~\cite{hoskins2019diagnostic,szabo2013diagnostic}.

\subsection{Related Work}

To avoid reliance on clean targets, recent work in natural images has explored novel self-supervised frameworks. Noise2Noise (N2N)~\cite{lehtinen2018noise2noise} showed that denoising models can be trained only using pairs of independently corrupted observations of the same scene, without clean targets. However, acquiring such paired data is often challenging in medical imaging. Especially in US imaging, speckle arises from deterministic wave interference and so remains stable under the same acquisition conditions, making it difficult to obtain independent noisy paired data for the same scenes~\cite{hoskins2019diagnostic,li2025speckle2self}.
To overcome the need for paired noisy data, blind-spot methods (BSN)~\cite{krull2019noise2void,lee2022ap} were proposed to learn from single noisy images by masking individual pixels and predicting them from their surroundings. While effective for spatially independent pixel-level noise, such pixel-wise masking is insufficient for US despeckling. Due to speckle's strong spatial dependency and clustered multi-pixel appearance, the masked pixel can still be inferred from neighboring correlated speckle patterns, leading the model to reproduce speckle rather than suppress it (case A in \cref{fig:denoising_typesPLUSnoisetypes}(b))~\cite{li2025speckle2self}.
Variants of BSN, such as StructN2V~\cite{broaddus2020removing}, extend this framework by introducing predefined spatial priors over noise dependencies and perform well for noise with fixed directional correlations. However, US speckle exhibits complex spatial dependencies~\cite{jensen1997field,treeby2010k} that are difficult to precisely capture with fixed spatial priors, limiting their applicability to US despeckling.
To specifically address US speckle, Speckle2Self~\cite{li2025speckle2self} introduced a US-tailored self-supervised framework based on scale-invariant priors. It departs from traditional pixel-wise assumptions and aligns more closely with a generative perspective.
While effective at reducing speckle, such generative methods may oversmooth fine anatomical structures, reflecting the fundamental trade-off between speckle suppression and detail preservation in the absence of clean targets.

Rather than treating US speckle as spatially independent noise or relying on generative synthesis to implicitly remove it, we revisit the self-supervised masking objective by asking not only what information should be hidden, but also what spatial unit should be reconstructed. Although some BSN variants optimize multiple blind spots in one pass~\cite{wang2022blind2unblind}, their prediction targets remain isolated pixels or predefined sparse patterns rather than a contiguous region covering a correlated speckle neighborhood. Mask2Restore instead treats each contiguous masked block as a single region-level reconstruction target and jointly restores its entire area, changing the task from sparse blind-spot estimation to region-level contextual inpainting. We therefore reformulate despeckling as \textbf{contextual inpainting} with block-wise masking (\cref{fig:denoising_typesPLUSnoisetypes}(b)); this region-level objective disrupts local speckle replication and enforces inpainting from surrounding context.
Because the visible context can still contain correlated speckle, we introduce \emph{cross-resolution context regularization} (CRCR), which enforces agreement after upsampling among predictions from a shared network at multiple input resolutions. Since the network's receptive field is fixed in input coordinates, it covers a larger spatial extent of the original image at lower resolutions, exposing broader anatomical context to help suppress speckle while preserving fine anatomy.
We conduct experiments on realistic simulated data with references and in vivo carotid data without references, demonstrating improved speckle--detail balance over representative self-supervised baselines. Additional evaluations on unseen fine-structure cases and downstream cardiac segmentation further support practical utility.

\noindent\textbf{Contributions:}
\begin{itemize}
\item We formulate self-supervised US despeckling as block-masked contextual inpainting, reconstructing contiguous regions instead of individual center pixels to disrupt local speckle replication.
\item We propose cross-resolution context regularization (CRCR), which uses shared multi-resolution predictions to reduce residual speckle bias while preserving fine anatomy.
\item Extensive evaluations on simulated and in vivo data, unseen fine structures, and downstream segmentation demonstrate a favorable balance between speckle suppression and detail preservation.
\end{itemize}

\section{Methodology}
\label{sec:method}

\begin{figure*}[t]
\centering
\includegraphics[width=0.82\textwidth]{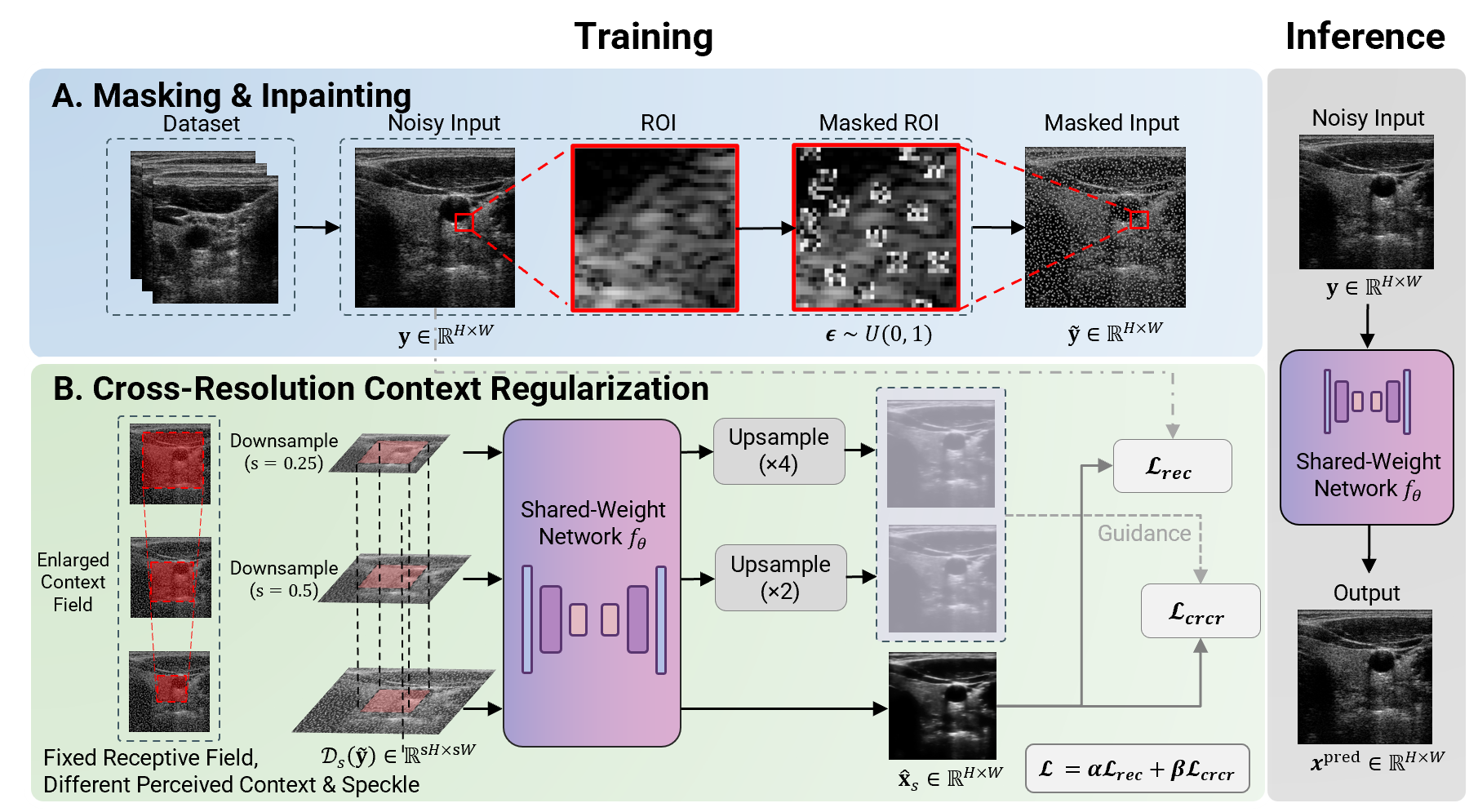}
\caption{An overview of the proposed framework.}
\label{fig:overview}
\end{figure*}

\subsection{Problem Statement}

Given a dataset consisting solely of noisy US images, our goal is to suppress speckle while preserving underlying fine structures as much as possible. Under linear propagation and weak-scattering assumptions, an RF-domain US image can be approximated as $\mathbf{r}=\mathbf{h}*\boldsymbol{\rho}$, where $\boldsymbol{\rho}$ denotes the tissue-scattering function, $\mathbf{h}$ is the imaging system's point spread function (PSF), and $*$ denotes convolution~\cite{jensen1997field,treeby2010k}. Because the PSF has nonzero spatial support, each scattering event contributes to a local neighborhood in the RF image. The resulting interference remains spatially correlated after envelope detection of the RF signal, causing a speckle grain to occupy multiple neighboring pixels rather than a single pixel.
Our method operates on these envelope US images. Let $\mathbf{y} \in \mathbb{R}^{H \times W}$ denote the observed envelope image, which is commonly represented in despeckling as $\mathbf{y}=\mathbf{x}\odot\mathbf{n}$, where $\mathbf{x}$ is the desired structure-preserving image and $\mathbf{n}$ is the multiplicative speckle component~\cite{yu2002speckle}. This compact representation does not imply that $\mathbf{n}$ is pixel-wise independent: its local correlation is inherited from the underlying RF-domain image formation. Consequently, pixel-wise blind-spot masking~\cite{krull2019noise2void} cannot fully exclude speckle information, because a masked pixel can still be inferred from neighboring pixels affected by the same local speckle pattern. We therefore learn a parameterized mapping $f_\theta: \mathbb{R}^{H \times W}\to\mathbb{R}^{H \times W}$ that estimates $\mathbf{x}$ from $\mathbf{y}$ while restricting access to these locally correlated speckle cues, as described next.

\subsection{Contextual Block-wise Masking}

Motivated by the multi-pixel spatial dependency of US speckle, we formulate despeckling as a block-wise masked inpainting problem, rather than relying on pixel-wise masking~\cite{krull2019noise2void} or scale-invariant priors~\cite{li2025speckle2self}. The distinction is not merely a larger blind spot around a center pixel: each sampled block defines one contiguous region-level target, whose entire area is reconstructed jointly from the visible context.
Specifically, we construct a binary mask $\mathbf{M} \in \{0, 1\}^{H \times W}$ by sampling anchor points at an anchor sampling ratio $r$ and centering an $N \times N$ square block at each selected point. The union of these square blocks defines the masked region, where $M_{\mathbf{u}}=0$ indicates a masked pixel at location $\mathbf{u}$. Thus, $r$ controls the anchor density rather than the final masked-pixel ratio, which also depends on $N$ and block overlap. Masked regions are then filled with random samples from a uniform distribution over the normalized intensity range.
\begin{equation}
\tilde{\mathbf{y}} = \mathbf{M} \odot \mathbf{y} + (\mathbf{1} - \mathbf{M}) \odot \boldsymbol{\epsilon}, \quad \boldsymbol{\epsilon} \sim \mathcal{U}(0, 1)
\end{equation}
Compared with zero filling, uniform random filling avoids introducing a visually trivial constant-valued hole and makes masked locations harder to identify from simple intensity cues. This discourages the network from exploiting the artificial mask pattern and encourages it to infer missing content from broader anatomical context.
By masking compact local regions rather than isolated pixels, this strategy reduces direct replication of local speckle patterns and forces the network to inpaint masked regions from the surrounding context.
However, although block-wise masking prevents direct speckle reproduction, the surrounding context still contains correlated speckle due to its high spatial dependency. As a result, masked inpainting alone may remain biased toward residual speckle patterns. This motivates the need to seek cues from a broader spatial context, where reconstruction can be guided more by tissue organization than by the local speckle pattern.

\subsection{Cross-Resolution Context Regularization}

To reduce residual speckle bias in context during inpainting, we introduce an implicit regularization, termed cross-resolution context regularization (CRCR).
Let $S = \{1, 0.5, 0.25\}$ denote the set of resolution factors. We define $\mathcal{D}_s: \mathbb{R}^{H \times W} \to \mathbb{R}^{sH \times sW}$ and $\mathcal{U}_s: \mathbb{R}^{sH \times sW} \to \mathbb{R}^{H \times W}$ as the spatial downsampling and upsampling operators, respectively.
The model's prediction $\hat{\mathbf{x}}_s \in \mathbb{R}^{H \times W}$ at scale
$s$ is given by
\begin{equation}
\hat{\mathbf{x}}_s = (\mathcal{U}_s \circ f_\theta \circ \mathcal{D}_s)(\tilde{\mathbf{y}})
\end{equation}
All resolution branches share the same full-resolution masked image $\tilde{\mathbf{y}}$ and therefore the same mask realization; $\mathcal{D}_s$ changes only its spatial resolution.
Downsampling modifies the input statistics in two complementary ways:
{\itshape 1)} Under a fixed receptive field of the same network, lower-resolution inputs correspond to larger physical receptive regions, enabling the network to rely on broader context rather than localized residual speckle patterns, as shown in \cref{fig:overview}. {\itshape 2)} Downsampling $\mathcal{D}_s$ tends to attenuate high-frequency components, weakening the relative influence of speckle compared to continuous tissue structures.
As a result, lower-resolution branches tend to focus on context-dominant predictions that are less affected by residual speckle bias.
These predictions serve only as consistency guidance rather than clean targets, because downsampling can also attenuate fine anatomical details. We enforce pairwise consistency among all scale predictions after upsampling:
\begin{equation}
\mathcal{L}_{crcr} = \frac{1}{2} \sum_{s \in S} \sum_{k \in S, k \neq s} \left\| \hat{\mathbf{x}}_s - \hat{\mathbf{x}}_k \right\|_1
\end{equation}
The CRCR term is optimized symmetrically across scales, whereas reconstruction supervision is applied only to the original-resolution branch ($s=1$). Their joint optimization therefore anchors the full-resolution prediction to the observed image while allowing the lower-resolution, broader-context predictions to regularize it. In this sense, the lower-resolution branches provide contextual guidance through the combined objective rather than through one-way supervision.

\subsection{Optimization and Training Process}

The training process is driven by a joint optimization objective that balances mask-induced inpainting and cross-resolution regularization. The total loss $\mathcal{L}_{\text{total}}$ is defined as:
\begin{equation}
\mathcal{L}_{\text{total}} = \alpha \mathcal{L}_{rec} + \beta \mathcal{L}_{crcr}
\end{equation}
The masked reconstruction loss ($\mathcal{L}_{rec}$) is applied within masked regions at full resolution ($s=1$):
\begin{equation}
\mathcal{L}_{rec} = \left\| (\mathbf{1}-\mathbf{M})\odot(\hat{\mathbf{x}}_1-\mathbf{y}) \right\|_2^2
\end{equation}
The reconstruction loss is evaluated only over the masked regions. As in masked self-supervised denoising~\cite{krull2019noise2void}, the observed image $\mathbf{y}$ provides the reconstruction target, while the corresponding target regions are absent from the masked input $\tilde{\mathbf{y}}$. The reconstruction term preserves image-specific structure and detail at full resolution, block-wise masking limits direct access to locally correlated speckle cues, and CRCR promotes agreement with predictions obtained from broader context. 
During inference, masking and multi-resolution branches are removed. The final despeckled output $\hat{\mathbf{x}}$ is obtained by directly applying $f_\theta$ to the original input $\mathbf{y}$ without additional inference cost.

\section{Experiments and Results}
\label{sec:experiments}

\subsection{Datasets and Experimental Setup}

We evaluate our method on multiple datasets spanning both simulated and in vivo US settings.
{\itshape a) Simulated US Dataset}~\cite{li2025speckle2self}: 538 simulated $512\!\times\!512$ envelope US images with clean targets generated by an acoustic-physics simulator, enabling PSNR/SSIM evaluation. The clean target corresponds to the ideal continuous-space scattering distribution before scatterer discretization and acoustic image formation. For evaluation, this distribution is represented on the same image grid and used as the idealized speckle-free structural output.
{\itshape b) Carotid US Dataset}~\cite{li2025speckle2self}: 523 carotid envelope US images ($512\!\times\!512$) without ground truth, used for qualitative and no-reference quantitative evaluation.
{\itshape c) Fine-Structure Generalization Dataset}: a simulated test set of 50 images under the same settings~\cite{li2025speckle2self}, embedded with small irregular structures unseen during training to assess generalization. These structures are introduced into the ideal continuous-space scattering distribution before scatterer discretization and US simulation, rather than being overlaid in the resulting image domain.
{\itshape d) CAMUS Dataset}~\cite{leclerc2019deep}: a public cardiac US dataset with expert annotations at end-diastolic (ED) and end-systolic (ES) frames, used to evaluate the impact of despeckling on downstream segmentation.
For the simulated and carotid datasets, we use the public train/test partitions released with the original dataset, without introducing an additional validation partition. Separate models are trained for the simulated and carotid datasets. The model evaluated on the fine-structure set is trained only on the original simulated training partition. For CAMUS, the learning-based despeckling models are retrained on the CAMUS training partition, and downstream evaluation uses the labeled ED and ES frames from 12 randomly selected patients in its held-out test partition.

\subsection{Baselines and Evaluation Metrics}

We compare against representative self-supervised baselines, including N2N~\cite{lehtinen2018noise2noise}, Noise2Void (N2V)~\cite{krull2019noise2void}, StructN2V~\cite{broaddus2020removing}, Deep Image Prior (DIP)~\cite{ulyanov2018deep}, AdaReNet~\cite{liu2025rotation}, Pixel2Pixel~\cite{ma2025pixel2pixel}, Blind2Unblind~\cite{wang2022blind2unblind}, and Speckle2Self~\cite{li2025speckle2self}, as well as the classical BM3D~\cite{dabov2007image} and NLM~\cite{coupe2009nonlocal} filters. Each learning-based baseline retains its original architecture and default optimization protocol; in particular, DIP and Pixel2Pixel use the stopping criteria and iteration schedules specified by their respective methods. BM3D and NLM are tuned only on the corresponding training split, after which their parameters are fixed for test evaluation. For N2N on the simulated dataset, each underlying scene includes two noisy observations: the default simulated image and a paired observation generated from the same scene after changing a subset of imaging parameters to produce a different speckle realization. N2N therefore has access to two noisy observations per scene, whereas each single-image learning method uses one. N2N is not applied to the in vivo datasets, where such pairs are unavailable.
For datasets with clean references, we report PSNR, SSIM, and LPIPS~\cite{zhang2018unreasonable} as our primary metrics, evaluating pixel fidelity, structural preservation, and natural perceptual quality directly against ground truth.
Clean targets are unavailable for in vivo US due to the intrinsic speckle formation mechanism, so we additionally report no-reference metrics for both datasets. We report Naturalness Image Quality Evaluator (NIQE)~\cite{mittal2012making}, a no-reference metric originally designed for natural images; as US differs substantially from natural images, its trends should be interpreted only as a coarse, secondary signal.

\begin{figure*}[!t]
\centering
\includegraphics[width=0.8\textwidth]{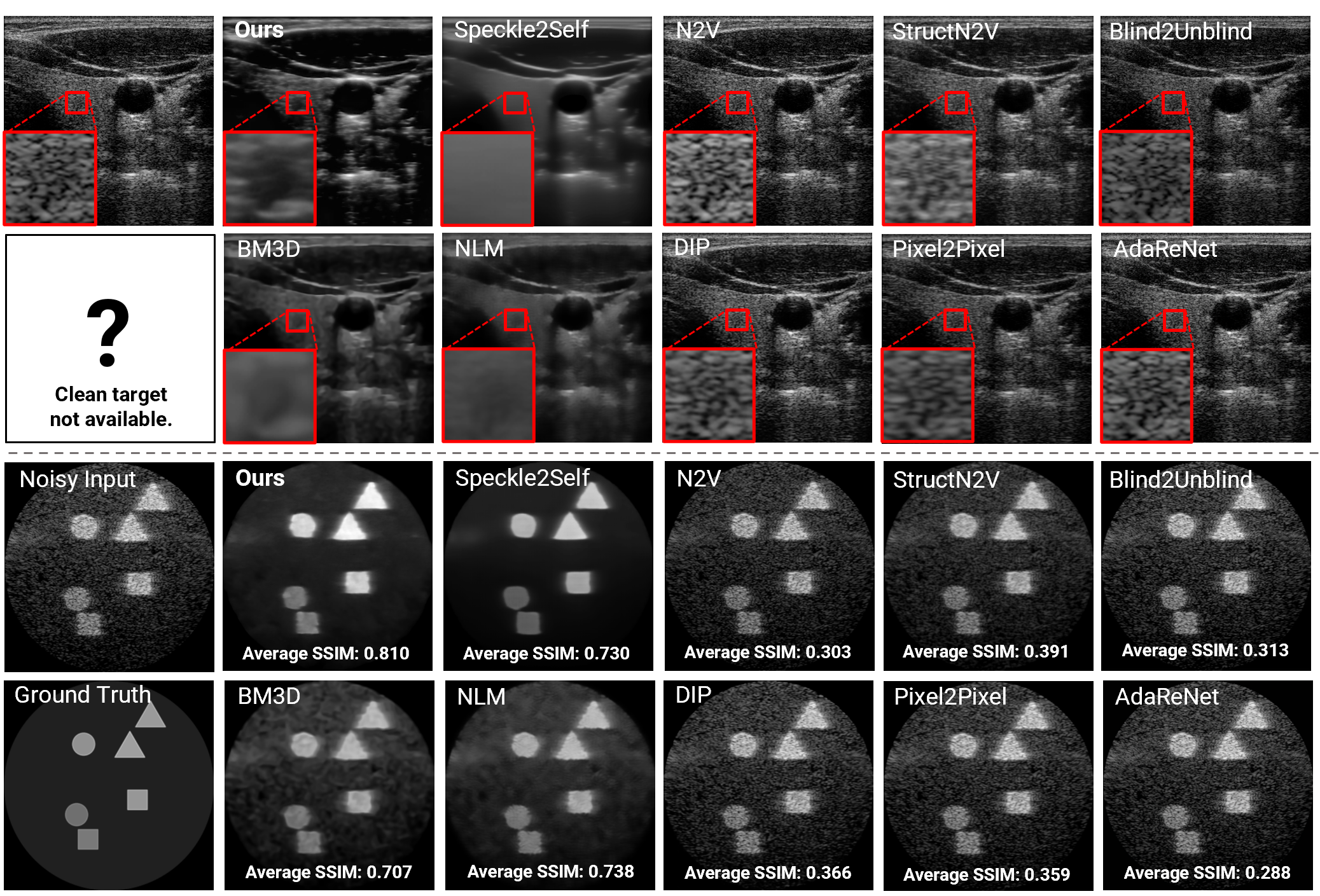}
\caption{Qualitative comparison on in vivo carotid and simulated datasets.}
\label{fig:comparision_all}
\end{figure*}

We further report the Edge Preservation Index (EPI)~\cite{jung2024unsupervised} and generalized Contrast-to-Noise Ratio (gCNR)~\cite{rodriguezmolares2020generalized}, two no-reference metrics tailored to US: EPI measures edge correlation before and after despeckling (closer to 1 is better), and gCNR measures the separability between signal and background intensity distributions within a region of interest. gCNR ranges from 0 to 1, with higher values indicating better target--background separability. For each in vivo test image, we select two ROIs and keep them fixed across all compared methods: a homogeneous thyroid-tissue region and a heterogeneous tissue region at a matched imaging depth. Without ground truth, each captures only a partial aspect of despeckling quality and can reward over-smoothed results that flatten texture without preserving structure; we therefore read EPI and gCNR jointly with visual inspection (\cref{fig:comparision_all}) and the unseen fine-structure evaluation (\cref{fig:abnormal_data}) rather than in isolation.

\subsection{Training Details}

Mask2Restore uses a standard U-Net~\cite{ronneberger2015u} backbone to ensure fair comparison with prior self-supervised denoising methods. For block-wise masking, the sampling ratio is set to $r{=}0.01$ and the block size to $N{=}7$. We set $\alpha{=}1$ and $\beta{=}0.1$, and train Mask2Restore in PyTorch for 200 epochs using Adam (learning rate 0.002, batch size 16) on a single NVIDIA RTX 5060 GPU. All learning-based methods use the same normalization and a common fixed random seed. The shared dataset-level augmentation consists of small-angle random rotations and horizontal flips; vertical flips are excluded because they invert the axial depth direction and are inconsistent with US acquisition geometry. Method-specific operations required by each baseline are otherwise retained.

\begin{table*}[!t]
\centering
\small
\caption{Quantitative comparison on simulated (with reference) and in vivo carotid (no reference) datasets. Best and second-best results are shown in \textbf{bold} and \underline{underline}. Reference-based metrics and NIQE are marked with $^{\mathrm{ref}}$ and $^{\mathrm{NR}}$ (no-reference), respectively.}
\label{tab:quantitative_comparison}
\resizebox{\textwidth}{!}{
\begin{tabular}{llccccccccc}
\toprule
\multirow{2}{*}{\textbf{Category}} & \multirow{2}{*}{\textbf{Method}} & \multicolumn{6}{c}{\textbf{Simulated (reference available)}} & \multicolumn{3}{c}{\textbf{Carotid (no reference)}} \\
\cmidrule(lr){3-8} \cmidrule(lr){9-11}
& & PSNR$^{\mathrm{ref}}\uparrow$ & SSIM$^{\mathrm{ref}}\uparrow$ & LPIPS$^{\mathrm{ref}}\downarrow$ & NIQE$^{\mathrm{NR}}\downarrow$ & EPI$\to\!1^\dagger$ & gCNR$\uparrow^\dagger$ & NIQE$^{\mathrm{NR}}\downarrow$ & EPI$\to\!1^\dagger$ & gCNR$\uparrow^\dagger$ \\
\midrule
\multirow{2}{*}{Non-learning} & BM3D~\cite{dabov2007image} & 17.83{\tiny$\pm$1.61} & 0.707{\tiny$\pm$0.042} & 0.496{\tiny$\pm$0.023} & 8.86{\tiny$\pm$0.15} & 1.66{\tiny$\pm$0.14} & 0.761{\tiny$\pm$0.200} & 8.86{\tiny$\pm$0.23} & \underline{0.68}{\tiny$\pm$0.16} & 0.774{\tiny$\pm$0.125} \\
& NLM~\cite{coupe2009nonlocal} & 17.92{\tiny$\pm$1.76} & \underline{0.738}{\tiny$\pm$0.032} & 0.455{\tiny$\pm$0.027} & 8.08{\tiny$\pm$0.26} & 1.70{\tiny$\pm$0.15} & 0.778{\tiny$\pm$0.207} & 7.32{\tiny$\pm$0.31} & 0.66{\tiny$\pm$0.16} & \underline{0.808}{\tiny$\pm$0.125} \\
\midrule
\begin{tabular}{@{}l@{}}Noisy--Noisy Pairs\end{tabular}

& Noise2Noise~\cite{lehtinen2018noise2noise} & 18.08{\tiny$\pm$1.72} & 0.679{\tiny$\pm$0.045} & 0.385{\tiny$\pm$0.011} & \textbf{7.20}{\tiny$\pm$0.24} & 1.90{\tiny$\pm$0.19} & 0.770{\tiny$\pm$0.221} & -- & -- & -- \\
\midrule
\multirow{8}{*}{\begin{tabular}{@{}l@{}}Single Noisy Images\end{tabular}}
& N2V~\cite{krull2019noise2void} & 17.38{\tiny$\pm$1.07} & 0.303{\tiny$\pm$0.009} & 0.618{\tiny$\pm$0.008} & 8.00{\tiny$\pm$0.45} & 12.41{\tiny$\pm$1.09} & 0.587{\tiny$\pm$0.236} & \underline{6.61}{\tiny$\pm$0.60} & 5.92{\tiny$\pm$1.71} & 0.382{\tiny$\pm$0.118} \\
& StructN2V~\cite{broaddus2020removing} & 17.89{\tiny$\pm$1.31} & 0.391{\tiny$\pm$0.018} & 0.598{\tiny$\pm$0.012} & 8.95{\tiny$\pm$0.50} & 8.26{\tiny$\pm$0.70} & 0.625{\tiny$\pm$0.246} & 9.49{\tiny$\pm$0.42} & 3.97{\tiny$\pm$1.09} & 0.454{\tiny$\pm$0.135} \\
& DIP~\cite{ulyanov2018deep} & 18.17{\tiny$\pm$1.08} & 0.366{\tiny$\pm$0.019} & 0.650{\tiny$\pm$0.019} & 8.43{\tiny$\pm$0.89} & 8.02{\tiny$\pm$0.63} & 0.613{\tiny$\pm$0.240} & 6.67{\tiny$\pm$0.70} & 3.09{\tiny$\pm$0.62} & 0.425{\tiny$\pm$0.131} \\

& AdaReNet~\cite{liu2025rotation} & 18.09{\tiny$\pm$0.87} & 0.288{\tiny$\pm$0.007} & 0.621{\tiny$\pm$0.006} & 7.74{\tiny$\pm$0.48} & 12.28{\tiny$\pm$1.21} & 0.580{\tiny$\pm$0.233} & 7.42{\tiny$\pm$0.53} & 4.60{\tiny$\pm$1.16} & 0.380{\tiny$\pm$0.120} \\
& Pixel2Pixel~\cite{ma2025pixel2pixel} & 18.45{\tiny$\pm$1.04} & 0.359{\tiny$\pm$0.013} & 0.646{\tiny$\pm$0.027} & 8.87{\tiny$\pm$0.75} & 8.25{\tiny$\pm$0.72} & 0.607{\tiny$\pm$0.242} & 8.03{\tiny$\pm$0.54} & 2.91{\tiny$\pm$0.68} & 0.445{\tiny$\pm$0.132} \\
& Blind2Unblind~\cite{wang2022blind2unblind} & 18.13{\tiny$\pm$0.89} & 0.313{\tiny$\pm$0.008} & 0.620{\tiny$\pm$0.007} & 8.82{\tiny$\pm$0.45} & 11.61{\tiny$\pm$1.08} & 0.581{\tiny$\pm$0.235} & 7.21{\tiny$\pm$0.62} & 4.97{\tiny$\pm$1.25} & 0.377{\tiny$\pm$0.117} \\
& Speckle2Self~\cite{li2025speckle2self} & \underline{19.27}{\tiny$\pm$0.87} & 0.730{\tiny$\pm$0.060} & \textbf{0.300}{\tiny$\pm$0.027} & 10.21{\tiny$\pm$0.54} & \textbf{1.10}{\tiny$\pm$0.11} & \textbf{0.819}{\tiny$\pm$0.218} & 8.79{\tiny$\pm$0.51} & 0.63{\tiny$\pm$0.17} & \textbf{0.838}{\tiny$\pm$0.120} \\
& \textbf{Ours} & \textbf{20.36}{\tiny$\pm$1.17} & \textbf{0.810}{\tiny$\pm$0.043} & \underline{0.319}{\tiny$\pm$0.024} & \underline{7.71}{\tiny$\pm$0.43} & \underline{1.27}{\tiny$\pm$0.15} & \underline{0.810}{\tiny$\pm$0.200} & \textbf{6.54}{\tiny$\pm$0.42} & \textbf{0.95}{\tiny$\pm$0.24} & 0.756{\tiny$\pm$0.122} \\
\bottomrule
\end{tabular}
}
\vspace{2pt}

{\footnotesize $^\dagger$EPI~\cite{jung2024unsupervised} and gCNR~\cite{rodriguezmolares2020generalized} are US-specific no-reference image quality metrics, unlike the general-purpose metrics reported alongside them.\par}
\end{table*}

\subsection{\texorpdfstring{Comparisons with the State of the Art}{Comparisons with the State of the Art}}

\begin{figure*}[t]
\centering
\includegraphics[width=0.8\textwidth]{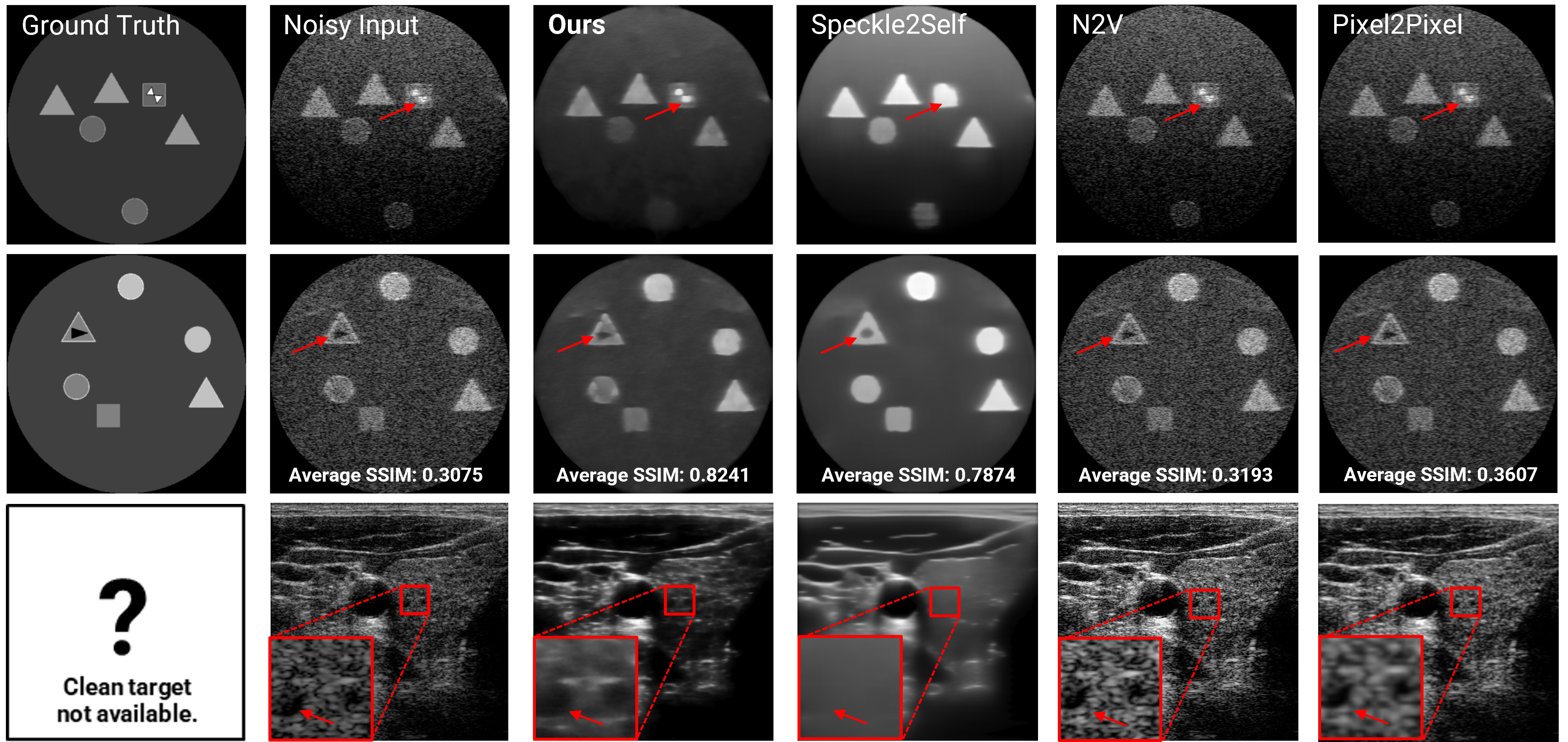}
\caption{Evaluation on unseen fine structures. The first two rows show simulated samples with atypical structures unseen during training, while the third row shows an in vivo carotid case. Red arrows mark fine structures preserved by our method but over-smoothed by competing methods.}
\label{fig:abnormal_data}
\end{figure*}

\Cref{tab:quantitative_comparison} summarizes the quantitative results, \cref{fig:comparision_all} provides qualitative comparisons on the standard test data, and \cref{fig:abnormal_data} further evaluates unseen fine structures. Our interpretation prioritizes reference-based structural metrics where clean targets are available and treats no-reference metrics as complementary evidence.

\textit{a) Simulated Dataset (with reference):}
Our method achieves the best PSNR (20.36) and SSIM (0.810), while remaining competitive in LPIPS. Among the reported metrics, SSIM provides the clearest quantitative evidence of structural preservation on this dataset because it directly compares local structure against the clean reference. Its advantage is also consistent across the two visual evaluations: our method better preserves anatomical boundaries in \cref{fig:comparision_all} and attains the highest SSIM (0.8241) while retaining previously unseen fine structures in \cref{fig:abnormal_data}. The limited SSIM gains of most generic self-supervised denoising baselines suggest a mismatch between their learning objectives and the multi-pixel spatial dependency of US speckle.

By contrast, NIQE provides limited task-specific discrimination. For example, Ours, AdaReNet, and N2V receive similar NIQE scores (7.71, 7.74, and 8.00) despite their substantially different SSIM values (0.810, 0.288, and 0.303) and visibly different speckle suppression in \cref{fig:comparision_all}; N2N obtains the best NIQE (7.20) but a lower SSIM (0.679). Speckle2Self obtains slightly better LPIPS and more favorable EPI and gCNR scores than our method, yet visibly attenuates contours and fine structures in \cref{fig:comparision_all,fig:abnormal_data}. This disagreement indicates that NIQE, EPI, and gCNR should be treated as auxiliary signals rather than decisive evidence of anatomical fidelity.

\textit{b) In Vivo Carotid Dataset (no reference):}
For real clinical data, reference-based metrics are not applicable, so NIQE, EPI, and gCNR are reported only as complementary indicators. N2N is not applicable because it requires independently corrupted pairs. NIQE again separates visually different outputs only weakly: Ours, N2V, and DIP obtain close scores of 6.54, 6.61, and 6.67, respectively, although N2V and DIP retain substantial speckle in \cref{fig:comparision_all}. Similarly, Speckle2Self achieves the highest gCNR but visibly attenuates fine structures in both \cref{fig:comparision_all} and the in vivo example in \cref{fig:abnormal_data}. Read jointly with the reference-based SSIM results and the unseen fine-structure test, the qualitative evidence supports that our method achieves a better speckle--detail trade-off.

\subsection{Generalization to Unseen Fine Structures}

To evaluate generalization to subtle structures unseen during training, we construct a 50-image simulated test set by embedding small atypical geometric patterns into simulated images. Each method is evaluated directly using its previously trained model: the simulation-trained model is used for the simulated cases, and the carotid-trained model is used for the in vivo case, without any fine-tuning. This setting tests robustness to complex unseen structures rather than adaptation to them. As shown in \cref{fig:abnormal_data}, while all methods improve SSIM over the noisy input, our method better preserves the embedded fine structures (red arrows), whereas competing methods, particularly Speckle2Self, tend to over-smooth them. Our method also achieves the highest SSIM (0.8241) on this set, quantitatively supporting its superior detail preservation in previously unseen cases. A similar trend appears in real carotid data (third row), where small hypoechoic regions are retained by our method but smoothed by competing approaches.

\subsection{Ablation Studies}
\label{sec:ablation}

We conduct two ablation studies. The first evaluates the contribution of the two proposed components. For the w/o Masking variant, the input is not corrupted ($\tilde{\mathbf y}=\mathbf y$), the reconstruction term becomes the full-image squared reconstruction error $\|\hat{\mathbf x}_1-\mathbf y\|_2^2$, and CRCR and all other settings remain unchanged. Using either block-wise masking or CRCR alone yields only limited gains (w/o Masking: 18.59\,dB / 0.372; w/o CRCR: 18.31\,dB / 0.383), whereas combining both improves performance substantially to 20.36\,dB / 0.810 (\cref{tab:ablation_study_component}). This supports our formulation: neither component alone is sufficient for effective speckle removal; block-wise masking reduces local speckle reproduction, while CRCR further suppresses residual speckle bias. The second ablation studies the masking hyperparameters by varying $r$ and $N$ (\cref{tab:ablation_study_mask}). A moderate setting ($r{=}0.01$, $N{=}7$) performs best (SSIM 0.8104), while overly large masks or sampling ratios reduce performance by removing too much valid context. This highlights the need to balance disruption of local speckle correlation against preservation of sufficient information for reliable inpainting.

\begin{table}[!b]
\centering
\small
\setlength{\tabcolsep}{5.5pt}
\caption{Component ablation. \checkmark\ indicates the component is enabled. Quantitative scores are computed on simulated data with references (below); in vivo examples are included only for qualitative comparison (above).}
\label{tab:ablation_study_component}
\begin{tabular}{lcc|cc}
\toprule
\multirow{2}{*}{\textbf{Variant}} & \multicolumn{2}{c|}{\textbf{Components}} & \multicolumn{2}{c}{\textbf{Metrics}} \\
\cmidrule(lr){2-3}\cmidrule(lr){4-5}
& \textbf{Masking} & \textbf{CRCR} & \textbf{PSNR (dB)$\uparrow$} & \textbf{SSIM$\uparrow$} \\
\midrule
w/o Masking & -- & \checkmark & 18.59 & 0.372 \\
w/o CRCR & \checkmark & -- & 18.31 & 0.383 \\
Full & \checkmark & \checkmark & \textbf{20.36} & \textbf{0.810} \\
\bottomrule
\end{tabular}
\par\vspace{0.9em}
\includegraphics[width=0.8\columnwidth]{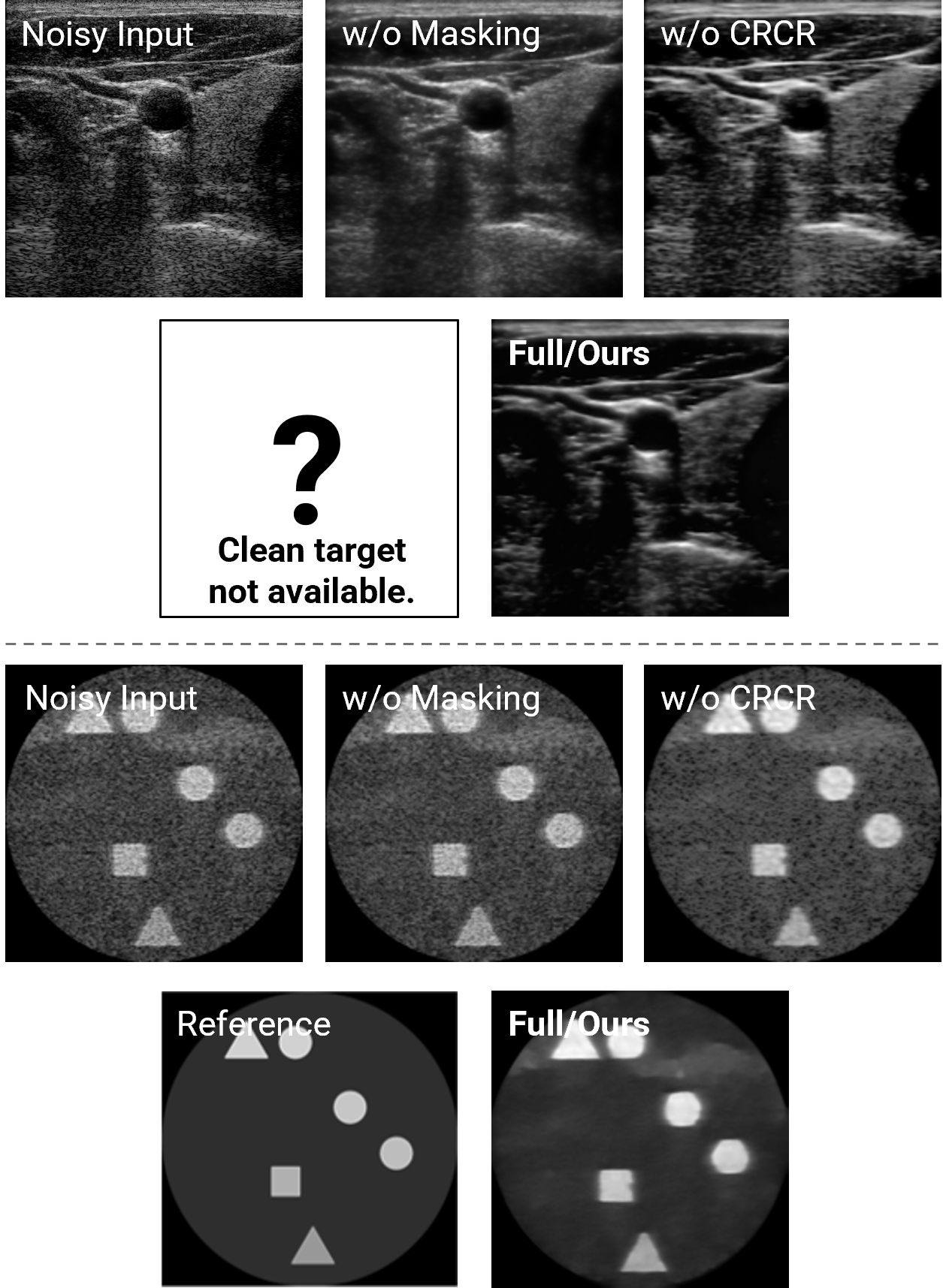}
\end{table}

\subsection{Downstream Segmentation Task}

We evaluate the downstream impact of despeckling via left atrium segmentation on the CAMUS cardiac US dataset using zero-shot SAM~2~\cite{ravi2024sam}. The learning-based despeckling methods are retrained on CAMUS rather than transferred from the carotid model. We evaluate 12 randomly selected patients from the held-out test partition and use only their labeled ED and ES frames for quantitative evaluation. After despeckling, each sequence is segmented by SAM~2 using prompts derived from the first labeled frame, with predictions propagated through the sequence. Dice and IoU are computed only at the labeled ED and ES frames. The results show that our method achieves the best performance (\cref{tab:final_segmentation_results}), suggesting better downstream segmentation from improved speckle suppression and structural preservation.

\begin{table}[!b]
\centering
\small
\begin{minipage}[t][0.50\textheight][s]{\columnwidth}
\centering
\caption{Sensitivity analysis of masking hyperparameters, measured by SSIM.}
\label{tab:ablation_study_mask}
\begin{tabular}{c|ccc}
\toprule
\multirow{2}{*}{Sampling ratio $r$} & \multicolumn{3}{c}{Block size $N$} \\
\cmidrule(lr){2-4}
 & 5 & 7 & 9 \\
\midrule
0.005 & 0.7594 & 0.7860 & 0.7970 \\
0.01  & 0.7532 & \textbf{0.8104} & 0.7284 \\
0.015 & 0.7496 & 0.6950 & 0.6993 \\
\bottomrule
\end{tabular}

\par\vfill
\caption{Downstream left atrium segmentation performance on CAMUS after despeckling, evaluated with identical SAM~2 prompts. Best and second-best results are shown in \textbf{bold} and \underline{underline}.}
\label{tab:final_segmentation_results}

\begin{tabular}{lcc}
\toprule
\textbf{Method} & \textbf{Dice ($\uparrow$)} & \textbf{IoU ($\uparrow$)} \\
\midrule
Noisy (Baseline) & 0.7664 & 0.6438 \\
\midrule
AdaReNet         & 0.7546 & 0.6314 \\
DIP              & 0.7686 & 0.6372 \\
Pixel2Pixel      & 0.7760 & 0.6484 \\
StructN2V        & 0.7782 & 0.6488 \\
N2V              & 0.7798 & 0.6641 \\
Speckle2Self     & 0.7865 & 0.6551 \\
BM3D             & 0.8098 & 0.6846 \\
NLM              & \underline{0.8145} & \underline{0.6899} \\
Blind2Unblind    & 0.7370 & 0.6087 \\
\midrule
\textbf{Ours}    & \textbf{0.8207} & \textbf{0.7002} \\
\bottomrule
\end{tabular}

\end{minipage}
\end{table}

\section{\texorpdfstring{Conclusion and Limitations}{Conclusion and Limitations}}
\label{sec:conclusion}

We presented a self-supervised framework for US despeckling that balances speckle suppression and fine-anatomy preservation without requiring clean targets. The method combines block-wise masked inpainting with cross-resolution context regularization (CRCR): block-wise masking reduces local speckle replication, while CRCR further suppresses residual speckle bias using broader, structure-dominant cues from lower-resolution branches. This yields anatomically coherent and speckle-suppressed reconstructions. Experiments on simulated and in vivo carotid US, together with evaluations on unseen fine structures and downstream cardiac segmentation, demonstrate its improved speckle--detail trade-off and practical value for subsequent image analysis.

In informal discussions, clinical collaborators consistently rated our despeckled outputs as an improvement over Speckle2Self; however, a subset preferred a lighter degree of despeckling, closer to the original speckled appearance. We believe this reflects a strong learned prior: sonographers are trained on speckle-containing images throughout clinical practice, and speckle is often treated as an intrinsic and expected visual feature of US rather than as pure noise to be removed.
This tension echoes a broader, longstanding debate in the despeckling literature between viewing speckle purely as noise and viewing it as a texture that can carry diagnostically relevant information (e.g., in tissue characterization and elastography)~\cite{damerjian2014speckle}. It suggests that automated image-quality metrics are, at best, imperfect proxies for clinical acceptance~\cite{huang2026defining} -- a concern reinforced by our own results, where even the US-specific no-reference metrics EPI and gCNR, not just general-purpose ones like PSNR, SSIM, or NIQE, can be inflated by over-smoothing (\cref{sec:experiments}) -- and that a formal reader study is required to properly assess diagnostic utility and adoption -- an important direction for future work.

\bibliographystyle{IEEEtran}
\bibliography{main}
\end{document}